# The Geometrization of Maxwell's Equations and the Emergence of Gravity and Antimatter

**Raymond J. Beach**

Lawrence Livermore National Laboratory, L-465, 7000 East Avenue, Livermore, CA 94551

E-mail: beach2@llnl.gov

ORCID 0000-0002-4099-130X

## ABSTRACT

Coupling the Maxwell tensor to the Riemann-Christoffel curvature tensor is shown to lead to a geometricized theory of electrodynamics. While this geometricized theory leads directly to the classical Maxwell equations, it also extends their physical interpretation by giving charge density, mass density and the four-velocity that describes their motions geometric definitions. Assuming the existence of a conserved energy-momentum tensor, all solutions to the geometricized theory of electrodynamics developed here are shown to be consistent with the emergence of gravity obeying the General Relativity field equation augmented by a term that mimics the properties of dark matter and/or dark energy. Finally, due to the symmetries of the theory, the properties and phenomenology of antimatter emerge in solutions.

## 1. INTRODUCTION

Electromagnetic and gravitational fields have long range interactions characterized by speed of light propagation; similarities that suggest these fields should be coupled together at the classical physics level. Although this coupling or unification is a well-worn problem with many potential solutions having been proposed,[1] it is fair to say that there is still no generally accepted classical field theory that can explain both electromagnetism and gravitation in a coupled or unified framework. Today, the descriptions of electromagnetic and gravitational fields are generally understood to be distinct and independent, with electromagnetic fields described by Maxwell's equations and gravitational fields described by Einstein's equation of General Relativity. The focus of this manuscript is the assessment of a recently proposed coupling between the Maxwell tensor and the Reimann-Christoffel curvature tensor that leads to a geometricized version of Maxwell's equations from which gravity and antimatter emerge. [2] A brief outline of the manuscript follows.

Assuming the geometry of nature is pseudo-Riemannian, the classical Maxwell equations will be shown to be a consequence of

$$F_{\mu\nu;\kappa} = a^{\lambda} R_{\lambda\kappa\mu\nu} \tag{1}$$

where $F_{\mu\nu}$ is the Maxwell tensor, $R_{\lambda\kappa\mu\nu}$ is the Riemann-Christoffel (R-C) curvature tensor, and $a^{\lambda}$ is a four-vector field related to the familiar vector potential of classical electromagnetism $A^{\lambda}$. After deriving Maxwell's equations from (1) mass is introduced using the conserved energy-momentum tensor for matter and electromagnetic fields,

$$\left(\rho_m u^{\mu} u^{\nu} + F^{\mu}{}_{\lambda} F^{\nu\lambda} - \frac{1}{4} g^{\mu\nu} F^{\rho\sigma} F_{\rho\sigma}\right)_{;\nu} = 0 \tag{2}$$

where $\rho_m$ is the scalar mass density, $u^{\lambda}$ is its four-velocity, and $g_{\mu\nu}$ the metric tensor. While Maxwell's equations are derivative to (1) alone, the inclusion of (2) leads to the Lorentz force law and the conservation of mass. Equations (1) and (2) together serve as the foundational axioms for the geometricized theory of electrodynamics that is developed within. One of the unusual features of the resulting classical field theory is that the properties and phenomenology for both matter and anti-matter emerge from it.

Beyond the framework (1) provides for the classical Maxwell equations, its coupling of the R-C tensor to the Maxwell tensor hint at the emergence of gravitational effects. The gravitational fields that do emerge in solutions to (1) and (2) do not necessarily satisfy Einstein's equation of General Relativity. Rather, as shown through the development within they are consistent with Einstein's equation of General Relativity augmented by an energy-momentum tensor that mimics the properties of dark matter and/or dark energy.

The goal of this manuscript is to show through an axiomatic development that a continuous field theory with (1) and (2) as axioms encompasses classical electrodynamics but then goes further with the emergence of gravitation and antimatter in its solutions. After developing the theory and discussing its various aspects, three specific solutions to (1) and (2) are presented. The purpose in developing these solutions is twofold: First, to provide a comparison of the solutions to (1) and (2) with the corresponding solutions of the classical Maxwell and Einstein Field Equations (M&EFEs), and second to demonstrate that the solutions to (1) and (2) go further than the classical M&EFEs by uniting electromagnetic and gravitational phenomena and also predicting the existence of antimatter. The first solution investigated is spherically symmetric and represents the asymptotic electric and gravitational fields of a non-rotating, charged particle. The second solution solves for radiative gravitational and electromagnetic waves with two distinct sub solutions, one with electromagnetic radiation in the presence of gravitational radiation, and the other with standalone gravitational radiation. This solution in particular illustrates the unification brought to electromagnetic and gravitational phenomena by (1). The third solution is for a maximally symmetric 3-dimensional subspace, for example, representing an isotropic and homogeneous universe.

Throughout the manuscript, geometric units will be used with a metric tensor having signature [+, +, +, -] in which spatial indices run from 1 to 3 with 4 the time index. The notation within uses commas before tensor indices to indicate ordinary derivatives and semicolons before tensor indices to indicate covariant derivatives. For the definitions of the R-C curvature tensor and the Ricci tensor, the conventions used by Weinberg are followed.[*]

## 2. GEOMETRIZED THEORY OF ELECTROMAGNETISM

### Maxwell's homogeneous equation and gauge invariance of $F_{\mu\nu}$

In this and the following section I give a short derivation of the classical Maxwell equations that follow from (1). The point in going through this purely formal development is to show that Maxwell's equations are derivative only to (1) and the algebraic properties of the R-C tensor with appropriate definitions for the charge density $\rho_c$ and the four-velocity $u^\nu$.

[*]See S. Weinberg, *Gravitation and Cosmology*, John Wiley & Sons, New York, NY 1972. The definition of the Riemann-Christoffel curvature tensor is $R^{\lambda}{}_{\mu\nu\kappa} \equiv \partial_{\kappa}\Gamma^{\lambda}{}_{\mu\nu} - \partial_{\nu}\Gamma^{\lambda}{}_{\mu\kappa} + \Gamma^{\eta}{}_{\mu\nu}\Gamma^{\lambda}{}_{\kappa\eta} - \Gamma^{\eta}{}_{\mu\kappa}\Gamma^{\lambda}{}_{\nu\eta}$ and the definition of the Ricci tensor is $R_{\mu\kappa} \equiv R^{\lambda}{}_{\mu\lambda\kappa}$.

Identifying $F_{\mu\nu}$ in (1) with the Maxwell tensor appears to be an attractive starting point in attempting to geometricize Maxwell's equations due to the following two indicial algebraic properties of the R-C tensor,

$$R_{\lambda\kappa\mu\nu} = -R_{\lambda\kappa\nu\mu} \tag{3}$$

and

$$R_{\lambda\kappa\mu\nu} + R_{\lambda\mu\nu\kappa} + R_{\lambda\nu\kappa\mu} = 0 \tag{4}$$

Contracting (3) with $a^\lambda$ gives by (1)

$$a^\lambda R_{\lambda\kappa\mu\nu} = -a^\lambda R_{\lambda\kappa\nu\mu} \quad \rightarrow \quad F_{\mu\nu;\kappa} = -F_{\nu\mu;\kappa} \tag{5}$$

and identifies $F_{\mu\nu}$ as antisymmetric,

$$F_{\mu\nu} = -F_{\nu\mu} \tag{6}$$

Contracting (4) with $a^\lambda$ gives by (1)

$$a^\lambda R_{\lambda\kappa\mu\nu} + a^\lambda R_{\lambda\mu\nu\kappa} + a^\lambda R_{\lambda\nu\kappa\mu} = 0 \quad \rightarrow \quad F_{\mu\nu;\kappa} + F_{\nu\kappa;\mu} + F_{\kappa\mu;\nu} = 0 \tag{7}$$

establishing the vanishing of the antisymmetrized covariant derivative of $F_{\mu\nu}$ which is also a statement of Maxwell's homogeneous equation,

$$F_{\mu\nu;\kappa} + F_{\nu\kappa;\mu} + F_{\kappa\mu;\nu} = 0 \quad \rightarrow \quad F_{\mu\nu,\kappa} + F_{\nu\kappa,\mu} + F_{\kappa\mu,\nu} = 0 \tag{8}$$

where the change from covariant to ordinary derivatives is justified by the affine connections in covariant derivatives that vanish on antisymmetrization.

While the foregoing derivation of Maxwell's homogeneous equation suggests that (1) is indeed an attractive starting point for geometricizing Maxwell's equations, one possible issue with (1) concerns the existence of solutions, *i.e.*, do they exist? I will come back to this question after first developing the full set of geometricized Maxwell's equations, but note in general that while solutions to the system of first order partial differential equations given by (1) for $F_{\mu\nu}$ do not exist for arbitrary $a^\lambda$ and $R_{\lambda\kappa\mu\nu}$, solutions do exist when $a^\lambda$ and $R_{\lambda\kappa\mu\nu}$ meet certain integrability conditions.

Having established the antisymmetry of $F_{\mu\nu}$ in (6), the vanishing of its antisymmetrized derivative in (7) or equivalently Maxwell's homogeneous equation in (8), and assuming the region under consideration is simply connected,[†] the converse to Poincaré's lemma states that $F_{\mu\nu}$ can itself be expressed as the anti-symmetrized derivative of a vector function,

$$F_{\mu\nu} = A_{\nu,\mu} - A_{\mu,\nu} \tag{9}$$

where $A_\mu$ is the classical electromagnetic vector potential. Equation (9) identifies $F_{\mu\nu}$ as gauge invariant, being unaffected when an arbitrary gradient field $\partial_\mu \varphi$ is added to $A_\mu$,

$$A_\mu \to A_\mu + \partial_\mu \varphi \tag{10}$$

**Maxwell's inhomogeneous equation and definitions for charge density and four-velocity**

Next, Maxwell's inhomogeneous equation and the definitions for the charge density $\rho_c$ and the four-velocity $u^\lambda$ forced by (1) are derived. Contracting the $\mu$ and $\kappa$ indices in (1) gives

$$F^{\mu\nu}{}_{;\mu} = a^\lambda R_{\lambda\mu}{}^{\mu\nu} = -a^\lambda R_\lambda{}^\nu \tag{11}$$

where $R_\lambda{}^\nu$ is the Ricci tensor. Defining the electromagnetic charge current density $\rho_c u^\nu$ by

$$\rho_c u^\nu \equiv a^\lambda R_\lambda{}^\nu \tag{12}$$

and then substituting $\rho_c u^\nu$ for $a^\lambda R_\lambda{}^\nu$ in (11) gives Maxwell's inhomogeneous equation

$$F^{\mu\nu}{}_{;\mu} = -\rho_c u^\nu \tag{13}$$

To establish the definitions that (12) imposes on the charge density $\rho_c$ and the four-velocity $u^\nu$, the following identity valid for any non-null four-vector $W^\mu$ is used,

$$W^\mu = \sqrt{\left|W^\rho W_\rho\right|}\,\frac{W^\mu}{\sqrt{\left|W^\sigma W_\sigma\right|}} \tag{14}$$

[†] Here simply connected means a region that can be continuously deformed to a point.

With the aid of (14) any $W^{\mu}$ satisfying $W^{\mu}W_{\mu} \neq 0$ can be recast as the product of a scalar density $\rho$ and a four-velocity $u^{\mu}$

$$W^{\mu} = \rho u^{\mu} \tag{15}$$

where the scalar density is defined by

$$\rho \equiv \pm\sqrt{\left|W^{\rho}W_{\rho}\right|} \tag{16}$$

and the four-velocity by

$$u^{\mu} \equiv \pm\frac{W^{\mu}}{\sqrt{\left|W^{\sigma}W_{\sigma}\right|}} \tag{17}$$

Note that (17) leads to different normalizations for the four-velocity $u^{\mu}$ depending on whether $W^{\mu}$ is time-like ( $W^{\mu}W_{\mu} < 0$ ) or space-like ( $W^{\mu}W_{\mu} > 0$ )

$$u^{\mu}u_{\mu} = \begin{cases} -1 \text{ if } W^{\mu} \text{ is time-like } \; W^{\mu}W_{\mu} < 0 \\ +1 \text{ if } W^{\mu} \text{ is space-like } \; W^{\mu}W_{\mu} > 0 \end{cases} \tag{18}$$

Time-like $W^{\mu}$ correspond to subluminal $u^{\mu}$ while space-like $W^{\mu}$ correspond to superluminal $u^{\mu}$. Identifying $W^{\nu}$ with $a^{\lambda}R_{\lambda}{}^{\nu}$ in equation (12) and using (16) and (17) now gives,

$$\rho_c \equiv \pm\sqrt{\left|a^{\rho}R_{\rho}{}^{\kappa}a^{\sigma}R_{\sigma\kappa}\right|} \tag{19}$$

and

$$u^{\nu} \equiv \pm\frac{a^{\lambda}R_{\lambda}{}^{\nu}}{\sqrt{\left|a^{\rho}R_{\rho}{}^{\kappa}a^{\sigma}R_{\sigma\kappa}\right|}} \tag{20}$$

Equations (19) and (20) emphasize the underlying geometric character imposed by (1) on Maxwell's equations with both the charge density $\rho_c$ and the four-velocity field $u^{\nu}$ defined in terms of the Ricci tensor $R_{\lambda}{}^{\nu}$ and the four-vector $a^{\lambda}$. This geometrization of $\rho_c$ and $u^{\nu}$ is reminiscent of classical General

Relativity's geometric interpretation of mass density $\rho_m$ in terms of the curvature scalar $R$ and hints at the emergence of gravity which will be developed subsequently.

In the development leading up to Maxwell's inhomogeneous equation (13) I have not imposed the usual restriction on the four-velocity $u^\lambda$ that it be subluminal. I drop this requirement because I am attempting to develop a theory that flows from (1) axiomatically and there is nothing *a priori* that requires that $a^\lambda R_\lambda{}^\nu$ be time-like. However, in the analysis that follows I will limit my focus to the case of subluminal velocities.

## Conservation of charge

The conservation of charge follows immediately from Maxwell's inhomogeneous equation (13) and the antisymmetry of $F_{\mu\nu}$. Taking the covariant divergence of Maxwell's inhomogeneous equation (13)

$$F^{\mu\nu}{}_{;\mu;\nu} = -\left(\rho_c u^\nu\right)_{;\nu} \tag{21}$$

and noting that $F^{\mu\nu}{}_{;\mu;\nu} \equiv 0$ , which is an identity for any antisymmetric tensor $F^{\mu\nu}$, establishes the conservation of charge

$$\left(\rho_c u^\nu\right)_{;\nu} = 0 \tag{22}$$

## Lorentz force law and conservation of mass

Continuing along the lines of (1) which was empirically chosen to reproduce the classical Maxwell equations, the conserved energy-momentum tensor given in (2) is now used to introduce mass density into the theoretical development. Using (2) and the already derived Maxwell's equations, here I derive the Lorentz force law and the conservation of mass equation. To start I distribute the covariant derivative in (2) which gives

$$\left(\rho_m u^\nu\right)_{;\nu} u^\mu + \rho_m u^\mu{}_{;\nu} u^\nu + F^{\mu\lambda} F^\nu{}_{\lambda;\nu} + F^{\mu\lambda}{}_{;\nu} F^\nu{}_\lambda - \frac{1}{2} g^{\mu\nu} F^{\rho\sigma} F_{\rho\sigma;\nu} = 0 \tag{23}$$

With some substitutions and rearrangements using Maxwell's homogeneous equation (8) and Maxwell's inhomogeneous equation (13), (23) can be re written as

$$\left(\rho_m u^\nu\right)_{;\nu} u^\mu + \rho_m u^\mu{}_{;\nu} u^\nu - \rho_c F^\mu{}_\lambda u^\lambda = 0 \tag{24}$$

Contracting (24) with $u_\mu$, the 2nd and 3rd terms on the LHS are zeroed due to the normalization of $u_\mu$ (18) and the antisymmetry of $F_{\mu\nu}$ (6), respectively, leaving

$$\left(\rho_m u^\nu\right)_{;\nu} = 0 \tag{25}$$

the conservation of mass equation. Using (25) to zero out the conservation of mass term in (24) then leaves the Lorentz force law

$$\rho_m \frac{Du^\mu}{D\tau} = \rho_c F^\mu{}_\lambda u^\lambda \tag{26}$$

where $\frac{Du^\mu}{D\tau} \equiv u^\mu{}_{;\nu} u^\nu$.

Analogous to the definitions developed for $\rho_c$ and $u^\nu$ in (19) and (20), respectively, (26) is now used to define the mass density $\rho_m$ in terms of the fields $a^\lambda$, $R_\lambda{}^\nu$, and $F_{\mu\nu}$. Solving equation (26) for $\rho_m$ and then substituting for $\rho_c$ and $u^\nu$ using their definitions gives

$$\rho_m \equiv \frac{F^\mu{}_\lambda\, \rho_c\, u^\lambda}{u^\mu{}_{;\nu} u^\nu} = \frac{F^\mu{}_\lambda a^\sigma R_\sigma{}^\lambda}{\left(\dfrac{a^\rho R_\rho{}^\mu}{\sqrt{\left|a^\gamma R_\gamma{}^\kappa a^\delta R_{\delta\kappa}\right|}}\right)_{;\nu}\left(\dfrac{a^\alpha R_\alpha{}^\nu}{\sqrt{\left|a^\beta R_\beta{}^\eta a^\zeta R_{\zeta\eta}\right|}}\right)} \quad \text{(no sum on } \mu\text{)} \tag{27}$$

Notable in the definitions for $\rho_c$, $u^\nu$ and $\rho_m$ given by (19), (20) and (27), respectively, are the interrelationships imposed amongst these quantities by the commonality of their dependencies on $a^\sigma$, $R_\sigma{}^\mu$ and $F^\mu{}_\lambda$. This is an important difference between their interpretations in classical electrodynamic theory versus that in the geometricized electrodynamics being developed here.

## Relationship of $a^\lambda$ to the classical electromagnetic vector potential $A^\lambda$

In the development so far, equation (1) and the vector field $a^\lambda$ that appears in it are the only truly new pieces of physics that have been introduced. However, it turns out that $a^\lambda$ is not entirely new, being related to the vector potential of classical electromagnetism $A^\lambda$. To see this, take the covariant derivative of both sides of (9)

$$F_{\mu\nu;\kappa} = A_{\nu;\mu;\kappa} - A_{\mu;\nu;\kappa} \tag{28}$$

and compare it to (1) rewritten as

$$F_{\mu\nu;\kappa} = a^{\lambda} R_{\lambda\kappa\mu\nu} = -a_{\kappa;\mu;\nu} + a_{\kappa;\nu;\mu} \tag{29}$$

where the RHS of (29) follows from the commutation property of covariant derivatives. Equating the RHS's of (28) and (29) gives

$$-a_{\kappa;\mu;\nu} + a_{\kappa;\nu;\mu} = A_{\nu;\mu;\kappa} - A_{\mu;\nu;\kappa} \tag{30}$$

establishing the connection between $a^{\lambda}$ and the vector potential of classical electromagnetism $A^{\lambda}$.

**Integrability condition that must be satisfied if equation (1) has solutions**

I now return to the question of the existence of solutions to (1) for $F_{\mu\nu}$. It is well known that mixed systems of first order partial differential equations such as equation (1) do not necessarily admit solutions.[3] As a familiar example, consider a vector field $S_{\mu}$ defined by parallel transport, $S_{\mu,\nu} - \Gamma^{\sigma}{}_{\mu\nu} S_{\sigma} = 0$. For this equation to have a self-consistent (single-valued) solution in some region, the change in $S_{\mu}$ after being transported around any closed path in that region must vanish. Carrying this exercise through, the condition that must be satisfied for a solution to exist is found to be $S_{\sigma} R^{\sigma}{}_{\mu\nu\rho} = 0$, which is equivalent to requiring the covariant derivatives of $S_{\mu}$ commute, *i.e.*, $S_{\mu;\nu;\rho} - S_{\mu;\rho;\nu} = 0$. The conclusion here is parallel transport can only be used to define a self-consistent vector field $S_{\mu}$ if the integrability condition $S_{\sigma} R^{\sigma}{}_{\mu\nu\rho} = 0$ is satisfied.[4]

An analogous argument to that just used to establish the necessary condition for a vector field $S_{\mu}$ defined by parallel transport to be self-consistent can also be used to establish the necessary condition for the antisymmetric tensor field $F_{\mu\nu}$ defined by (1) to be self-consistent. Begin by calculating the change in $F_{\mu\nu}$ when it is transported around any closed path using (1). The integrability condition is then defined by forcing the change in $F_{\mu\nu}$ to vanish. Working through the details is straight forward and leads to the following condition that must be satisfied if the round-trip change in $F_{\mu\nu}$ is to vanish,

$$\left(a^{\rho} R_{\rho\kappa\mu\nu}\right)_{;\lambda} - \left(a^{\rho} R_{\rho\lambda\mu\nu}\right)_{;k} = -F_{\mu\sigma} R^{\sigma}{}_{\nu k\lambda} - F_{\sigma\nu} R^{\sigma}{}_{\mu\kappa\lambda} \tag{31}$$

With (31) as integrability condition that must be satisfied for equation (1) to be solvable, the question that naturally arises is this: Are these integrability conditions so restrictive that perhaps no solution to the proposed theory exists? However, as will be presented in section 3 solutions to (1) have been found.

**The derivation of the classical Maxwell's equations vs those derived using $F_{\mu\nu;\kappa} = a^{\lambda}R_{\lambda\kappa\mu\nu}$**

Except for the geometricized definitions of $\rho_c$ and $u^{\lambda}$ given by (19) and (20), respectively, Maxwell's equations derived using (1) are identical to those arrived at using the conventional Lagrangian-based classical derivation. This is surprising given these derivations have very different starting points.

In the classical derivation,[5] Maxwell's homogeneous equation must be taken as an axiom so that the vector potential $A_{\mu}$ can be introduced; $A_{\mu}$ being necessary for the definition of the matter-field interaction term $\rho_c A_{\lambda} u^{\lambda}$ in the action $I_M$ defined by

$$I_M = \int d^4x\sqrt{g(x)}\left(-\rho_m\sqrt{\left|u^{\lambda}u_{\lambda}\right|} + \rho_c A_{\lambda}u^{\lambda} - \frac{1}{4}F^{\rho\sigma}F_{\rho\sigma}\right) \tag{32}$$

where $g(x) = -\det(g_{\mu\nu})$. Maxwell's inhomogeneous equation and the Lorentz force law are then derived using a stationary-action calculation in which $A_{\mu}$ and the spatial positions of masses and charges are treated as the dynamic variables. Finally, an energy-momentum tensor, the same one as in equation (2) is derived as the functional derivative of the scalar $I_M$ (32) with respect to $g_{\mu\nu}$ and then shown to be conserved due to the general covariance of $I_M$.

In the derivation followed here, equation (1) is taken as an axiom and the Maxwell equations are derived using a 2-step development

$$F_{\mu\nu;\kappa} = a^{\lambda}R_{\lambda\kappa\mu\nu} \;\rightarrow\; \begin{cases} F_{\mu\nu;\kappa} + F_{\nu\kappa;\mu} + F_{\kappa\mu;\nu} = 0 \\ F^{\mu\nu}{}_{;\mu} = -a^{\lambda}R_{\lambda}{}^{\nu} \end{cases} \;\rightarrow\; \begin{cases} F_{\mu\nu,\kappa} + F_{\nu\kappa,\mu} + F_{\kappa\mu,\nu} = 0 \\ F^{\mu\nu}{}_{;\mu} = -\rho_c u^{\nu} \end{cases} \tag{33}$$

in which the first step depends only on (1) and the properties of the R-C tensor given in (3) and (4), and the second step on the definition of the charge current density given in (12). The Maxwell equations are identical regardless of the derivation used, and it is only the starting point axioms that are different. In the classical Lagrangian-based derivation a specific Lagrangian and Maxwell's homogeneous equation are taken as the starting point axioms while in the derivation followed here (1) and (2) are taken as the starting point axioms. Including the conserved energy-momentum tensor (2) as an axiom completes the derivation of the geometricized theory of electrodynamics with the conservation of mass (25) and the Lorentz force law (26) being derived from it and the already derived Maxwell's equations.

As an alternative to the derivation here in which (1) and (2) are taken as axioms, an equivalent set of axioms consists of (1) and the classical Lagrangian development using the action defined in (32). In this approach the conserved energy-momentum tensor given in (2) is dropped in favor of the Lagrangian development. This is justified because (2) is a consequence of the Lagrangian development and because Maxwell's homogeneous equation which is needed for the Lagrangian development is established using (1). Whether taking (1) and (2) as axioms, or taking (1) and the classical Lagrangian development as axioms, the resulting theories are identical with the same geometric definitions imposed on $\rho_m$, $\rho_c$ and $u^\lambda$, thus emphasizing again that the only new physics being proposed within is contained in equation (1).

One peculiar feature of the development shown in (33) is that equation (1) on its LHS exhibits an explicit dependence on both $a^\lambda$ and $R_{\lambda\kappa\mu\nu}$ while the classical Maxwell equations on the RHS contain neither $a^\lambda$ nor $R_{\lambda\kappa\mu\nu}$ explicitly. It is as if the classical Maxwell equations have conspired to hide any hint of their dependence on the new physics contained in (1). For example, with Maxwell's homogeneous equation any hint of curved space-time is hidden because the affine connections that appear in covariant derivatives vanish on antisymmetrization making this equation identical in curved and flat space-time. However, for Maxwell's inhomogeneous equation the situation is different. The charge current density $\rho_c u^\nu$ has a geometric underpinning given by (12) which vanishes in the absence of curvature. Contrary to this picture is the classical interpretation in which the charge current density is assumed to exist in flat space-time, making the classical interpretation of Maxwell's inhomogeneous equation at best an approximation to the geometricized version of it. Evident from the forgoing discussion is that deviations between the geometricized theory developed here and the classical Maxwell theory would be expected to be the greatest in those regions containing charge density.

## Emergence of gravity

The preceding discussion established that the equations of classical electromagnetism follow directly from (1) and (2) with appropriate definitions adopted for $\rho_m$, $\rho_c$ and $u^\lambda$. The question that naturally arises now is this, what does equation (1) have to say about gravity? Due to the coupling of the R-C tensor to the Maxwell tensor some form of gravity is expected to emerge in the solutions of (1) and (2). If these solutions are to have any physical relevance then the behavior of the emergent gravity will need to be close to the predictions of Einstein's General Relativity

$$G^{\mu\nu} = -8\pi\, T^{\mu\nu} \tag{34}$$

where $G^{\mu\nu} \equiv R^{\mu\nu} - \frac{1}{2} g^{\mu\nu} R$ is the Einstein tensor and $T^{\mu\nu}$ is the conserved energy-momentum tensor given in (2). Using the specific example of a spherically symmetric, non-rotating, charged particle that is given in section 3, the Reissner-Nordström metric is shown to be an exact solution to (1) and (2) establishing that the emergent gravity in the proposed theory is the same as that predicted by classical General Relativity (34) in the case of spherical symmetry. However, to determine if Einstein's field equation is a derivable consequence of (1) and (2), one must go further. To investigate this issue consider equation (2), the conserved energy-momentum tensor $T^{\mu\nu}{}_{;\nu} = 0$. An immediate consequence of $G^{\mu\nu}$ being independently conserved due to the Bianchi identity is that for any constant $\alpha$, one can define a tensor field $\Lambda^{\mu\nu}$ by

$$\Lambda^{\mu\nu} \equiv G^{\mu\nu} - \alpha T^{\mu\nu} \tag{35}$$

that is both symmetric,

$$\Lambda^{\mu\nu} = \Lambda^{\nu\mu} \tag{36}$$

and conserved,

$$\Lambda^{\mu\nu}{}_{;\nu} = 0 \tag{37}$$

The value of the constant $\alpha$ in (35) is completely arbitrary and without physical significance because $\Lambda^{\mu\nu}$ as defined can absorb any change in $\alpha$ such that (35) remains satisfied. Taking advantage of this arbitrariness and setting the value of the constant $\alpha = -8\pi$ then gives with a slight rearrangement of (35)

$$G^{\mu\nu} = -8\pi\, T^{\mu\nu} + \Lambda^{\mu\nu} \tag{38}$$

which is recognized as Einstein's equation of General Relativity (34) augmented on its RHS by the term $\Lambda^{\mu\nu}$. The $\Lambda^{\mu\nu}$ term in (38) exhibits the properties of an energy-momentum tensor appropriate for dark matter and/or dark energy, *viz.*, it is a conserved and symmetric tensor field, it is a source of gravitational fields in addition to energy-momentum tensor $T^{\mu\nu}$ for normal matter and normal energy and has no interaction signature beyond the gravitational fields it sources.

At this point it is important to recognize that (38) is a trivial result with no physical significance in the theory being proposed here. This follows because any solution of (1) and (2) must necessarily be a solution of (38) for some choice $\Lambda^{\mu\nu}$. In fact, the validity of (38) rests only on the existence of a conserved energy-momentum tensor and the properties of the R-C tensor, and so will be true in any physical theory having a conserved energy-momentum tensor. However, the interesting point in the context of the proposed theory here is that the value of $\Lambda^{\mu\nu}$ can be calculated from solutions of (1) and (2) without postulating the existence of dark matter and/or dark energy.

In summary, gravitation does emerge as a manifestation of the geometricized theory of electrodynamics based on equations (1) and (2). Specifically, it is the coupling of the derivatives of the Maxwell tensor to the R-C tensor in (1) that brings gravitation into the picture. Importantly, the gravitational theory that emerges does not obey the classical General Relativity field equation (34), although any solution of (1) and (2) must necessarily be a solution of (38) for some choice of $\Lambda^{\mu\nu}$. While viewing gravitation as a manifestation of electromagnetism and vice versa is not new [6, 7, 8, 9, 10], the specific approach being followed here with equation (1) is.

**Global symmetries of equations (1) and (2)**

Table I lists the six fields that have been used in the development of the theory that flows from equations (1) and (2). Based on these developments, the fields fall into two categories: The fundamental fields $a_\lambda$, $g_{\mu\nu}$ and $F_{\mu\nu}$ that (1) and (2) solve for, and the remaining fields $\rho_c$, $u^\lambda$ and $\rho_m$ which are defined in terms of the fundamental fields by (19), (20), and (27), respectively.

**Table I. Fields**

| Field | Description |
| --- | --- |
| $a_\lambda$ | Four-vector coupling gravitation and electromagnetism |
| $g_{\mu\nu}$ | Metric tensor |
| $F_{\mu\nu}$ | Maxwell tensor |
| $\rho_c$ | Charge density scalar field – defined by equation (19) |
| $u^\lambda$ | Four-velocity vector field – defined by equation (20) |
| $\rho_m$ | Mass density scalar field – defined by equation (27) |

To help identify the global symmetries of (1) and (2), these equations are collected here along with equations (12) and (27) which give the definitions of the charge current density and mass density, respectively:

$$F_{\mu\nu;\kappa} = a^{\lambda} R_{\lambda\kappa\mu\nu} \tag{1}$$

$$\left( \rho_m u^{\mu} u^{\nu} + F^{\mu}{}_{\lambda} F^{\nu\lambda} - \frac{1}{4} g^{\mu\nu} F^{\rho\sigma} F_{\rho\sigma} \right)_{;\nu} = 0 \tag{2}$$

$$a^{\lambda} R_{\lambda}{}^{\nu} \equiv \rho_c u^{\nu} \tag{12}$$

$$\rho_m \equiv \frac{F^{\mu}{}_{\lambda} a^{\sigma} R_{\sigma}{}^{\lambda}}{\left( \dfrac{a^{\rho} R_{\rho}{}^{\mu}}{\sqrt{\left| a^{\gamma} R_{\gamma}{}^{\kappa} a^{\delta} R_{\delta\kappa} \right|}} \right)_{;\nu} \left( \dfrac{a^{\alpha} R_{\alpha}{}^{\nu}}{\sqrt{\left| a^{\beta} R_{\beta}{}^{\eta} a^{\zeta} R_{\zeta\eta} \right|}} \right)} \quad \text{(no sum on } \mu) \tag{27}$$

These four equations illustrate the three global symmetries imposed on the fields listed in Table I. The first of these global symmetries corresponds to charge-conjugation,

$$\begin{pmatrix} a^{\lambda} \\ g_{\mu\nu} \\ F^{\mu\nu} \\ \rho_c \\ u^{\lambda} \\ \rho_m \end{pmatrix} \to \begin{pmatrix} -a^{\lambda} \\ g_{\mu\nu} \\ -F^{\mu\nu} \\ -\rho_c \\ u^{\lambda} \\ \rho_m \end{pmatrix} \tag{39}$$

the second corresponds to matter–antimatter conjugation,

$$\begin{pmatrix} a^{\lambda} \\ g_{\mu\nu} \\ F^{\mu\nu} \\ \rho_c \\ u^{\lambda} \\ \rho_m \end{pmatrix} \to \begin{pmatrix} -a^{\lambda} \\ g_{\mu\nu} \\ -F^{\mu\nu} \\ \rho_c \\ -u^{\lambda} \\ \rho_m \end{pmatrix} \tag{40}$$

as will be discussed in the next section, and the third to the product of the first two,

$$\begin{pmatrix} a^{\lambda} \\ g_{\mu\nu} \\ F^{\mu\nu} \\ \rho_c \\ u^{\lambda} \\ \rho_m \end{pmatrix} \to \begin{pmatrix} a^{\lambda} \\ g_{\mu\nu} \\ F^{\mu\nu} \\ -\rho_c \\ -u^{\lambda} \\ \rho_m \end{pmatrix} \tag{41}$$

All three transformations leave equations (1), (2), (12) and (27) unchanged. Adding the identity transformation to these symmetries forms the Klein-4 group, with the product of any two of the symmetries (39) through (41) giving the remaining symmetry.

Note that among the fields of the theory only $g_{\mu\nu}$ and $\rho_m$ are unchanged by the symmetry transformations, a fact that will be useful for defining self-consistency equations that lead to a mechanism for quantizing the mass, charge, and angular momentum of the particle-like solutions as discussed in section 3. Finally, in addition to general covariance and the global symmetries of (39) through (41), the proposed theory exhibits the electromagnetic gauge invariance of classical electromagnetism as detailed in (9) and (10).

**Emergence of antimatter**

One of the interesting features of equations (1) and (2) is that the properties of antimatter emerge naturally in their solutions. Traditionally, these properties emerge in quantum mechanical treatments but here emerge in the context of a classical continuous field theory due to the global symmetry (40) exhibited by (1) and (2) with every matter containing solution having a corresponding antimatter solution generated by the transformation (40). To demonstrate that transformation (40) indeed conforms to the expected properties of matter–antimatter conjugation, here I will examine its impact on a test particle in an external electromagnetic and gravitational field and show that the resulting phenomenology corresponds to that of matter and antimatter.

To begin, consider the action of the symmetry transformation (40) on the four-velocity, transforming $u^{\lambda} \to -u^{\lambda}$, which is physically equivalent to the view that a particle's antiparticle is the particle moving backwards through time.[11] By convention I take matter to have a positive time component for its four-velocity and antimatter to have a negative time component

$$u^4 \begin{cases} > 0 \text{ for matter} \\ < 0 \text{ for antimatter} \end{cases} \tag{42}$$

To illustrate the behavior of matter and antimatter in an external electromagnetic or gravitational field it will be convenient to work in a locally inertial frame and express the test particle's $u^\lambda$ as

$$u^\lambda = \frac{dx^\lambda}{d\tau} \tag{43}$$

where $d\tau$ is the proper time. The symmetry transformation (40) with $u^\lambda \to -u^\lambda$ is then equivalent to $d\tau \to -d\tau$ in (43). In a locally inertial coordinate system, this motivates the following expression for the four-velocity in terms of the coordinate time $t$,

$$u^\lambda = \frac{dx^\lambda}{d\tau} = s_{m-a}\,\gamma \frac{dx^\lambda}{dt} = s_{m-a}\,\gamma \begin{pmatrix} \vec{v} \\ 1 \end{pmatrix} \tag{44}$$

where $\vec{v}$ is the ordinary 3-space velocity of the charge and mass density, $\gamma = 1/\sqrt{1-\vec{v}^{\,2}}$, and $s_{m-a}$ is a parameter defined by

$$s_{m-a} = \begin{cases} +1 \text{ for matter} \\ -1 \text{ for antimatter} \end{cases} \tag{45}$$

Note that in transforming a particle to its antiparticle (40) does not change the sign of the charge density $\rho_c$. This seems contradictory to the view today in which the charge of an antiparticle is opposite to that of its associated particle. However, because the external fields acting on the particles are not changed in the transformation of particle to antiparticle, transformation (40) leads to the expected behavior, with the antiparticle appearing to have the opposite charge density as that of its associated particle. To see this consider a region with an external electromagnetic field defined by

$$F_{\mu\nu} = \begin{pmatrix} 0 & B_z & -B_y & E_x \\ -B_z & 0 & B_x & E_y \\ B_y & -B_x & 0 & E_z \\ -E_x & -E_y & -E_z & 0 \end{pmatrix} \tag{46}$$

Starting with the Lorentz force law (26) and then using (44) for the definition of $u^\lambda$ gives the following development

$$\rho_p \frac{Du^\mu}{D\tau} = \rho_c F^\mu{}_\lambda u^\lambda$$
$$\downarrow$$
$$\rho_p s_{m-a} \gamma \frac{d}{dt}\begin{pmatrix} s_{m-a} \gamma \vec{v} \\ s_{m-a} \gamma \end{pmatrix} = \rho_c \begin{pmatrix} 0 & B_z & -B_y & E_x \\ -B_z & 0 & B_x & E_y \\ B_y & -B_x & 0 & E_z \\ E_x & E_y & E_z & 0 \end{pmatrix}\begin{pmatrix} s_{m-a} \gamma v_x \\ s_{m-a} \gamma v_y \\ s_{m-a} \gamma v_z \\ s_{m-a} \gamma \end{pmatrix} \tag{47}$$
$$\downarrow$$
$$\rho_p \frac{d}{dt}\begin{pmatrix} \gamma \vec{v} \\ \gamma \end{pmatrix} = s_{m-a} \rho_c \begin{pmatrix} \vec{E} + \vec{v} \times \vec{B} \\ \vec{v} \cdot \vec{E} \end{pmatrix}$$

which on the last line ends up at the conventional form of the Lorentz force law except for the factor of $s_{m-a}$ on the RHS. This factor of $s_{m-a}$ gives the product $s_{m-a} \rho_c$ the appearance that antimatter charge density has the opposite sign to that matter charge density when interacting with an external electromagnetic field and so is consistent with the view that the charge of an antiparticle is opposite to that of its associated particle when interacting with an external electromagnetic field.

Next, I investigate the behavior of antimatter in an external gravitational field. There is no question about the gravitational fields generated by matter and antimatter, they are identical under the matter-antimatter symmetry (40), as $g_{\mu\nu} \to g_{\mu\nu}$. To understand whether antimatter is attracted or repelled by an external gravitational field, I again go to the Lorentz force law (26) but this time assume there is no electromagnetic field present, just an external gravitational field given by a Schwarzschild metric and generated by a central mass $m > 0$ that is composed of either matter or antimatter. The use of the Schwarzschild metric will be justified in Section 3 where a spherically symmetric solution for a point particle is developed. I explicitly call out $m > 0$ because I am endeavoring to develop a physical theory that flows axiomatically from equations (1) and (2), and at this point in the development there is nothing to preclude the existence of negative mass density $\rho_m < 0$, a consideration I will return to later. Placing a test particle having mass $m_{test}$ composed of either matter or antimatter a distance $r$ from the center of the gravitational field and assuming the test particle is initially at rest, the trajectory of the test particle is that of a geodesic

$$\frac{Du^\mu}{D\tau}=0$$
$$\downarrow$$
$$s_{m-a}\,\gamma\frac{du^\mu}{dt}=-\Gamma^\mu{}_{\nu\rho}u^\nu u^\rho \tag{48}$$
$$\downarrow$$
$$s_{m-a}\,\gamma\frac{d}{dt}\left(s_{m-a}\gamma\frac{d}{dt}\begin{pmatrix}r\\ \theta\\ \phi\\ t\end{pmatrix}\right)=-\Gamma^\mu{}_{\nu\rho}u^\nu u^\rho\approx-\Gamma^\mu{}_{44}u^4u^4=-\begin{pmatrix}\left(1-\frac{2m}{r}\right)\frac{m}{r^2}\\ 0\\ 0\\ 0\end{pmatrix}\left(\frac{s_{m-a}}{\sqrt{1-\frac{2m}{r}}}\right)^2=-\begin{pmatrix}\frac{m}{r^2}\\ 0\\ 0\\ 0\end{pmatrix}s_{m-a}{}^2$$

where $s_{m-a}=\pm 1$ as defined in (45) and references whether the test particle is composed of matter or antimatter. In (48) I have approximated the RHS using the initial at rest value of the test particle's four-velocity $u^\mu=\left(0,0,0,s_{m-a}/\sqrt{1-2m/r}\right)$, and additionally used the fact that the only nonzero $\Gamma^\mu{}_{44}$ in a Schwarzschild metric is $\Gamma^1{}_{44}=\left(1-\frac{2m}{r}\right)m/r^2$. Simplifying the LHS of the last line in (48) by noting that initially $\gamma=1$ gives

$$\frac{d^2r}{dt^2}\approx-\frac{m}{r^2} \tag{49}$$

which is independent of $s_{m-a}$, and so demonstrates that the proposed theory predicts both matter and antimatter test particles will be attracted by the source of the gravitational field, and this regardless of whether the source of the gravitational field is matter or antimatter. The result that the test particle is attracted toward the source of the gravitational field is also independent of whether the test particle's mass, $m_{test}$, is positive or negative, this because the geodesic trajectory (48) is independent of $m_{test}$.

As already noted, there appears to be nothing in equations (1) and (2) that precludes the possibility of negative mass density $\rho_m<0$. The existence of negative mass density is equivalent to the existence of antigravity because negative mass density generates gravitational fields that are repulsive, *viz.*, (49) with $m<0$. However, digging a little deeper, logical inconsistencies within the construct of the proposed theory do arise if negative mass density were to exist. As just shown, (49) with $m>0$ predicts a test particle at some distance from the origin will feel an attractive gravitational force regardless of whether the test particle is comprised of matter or antimatter and regardless of whether its mass is positive or negative. Now consider equation (49) with the central mass $m<0$. Using the same argument as in the previous section, the test particle in this case will feel a repulsive gravitational force regardless of whether it is comprised of matter or antimatter and regardless of whether its mass is positive or negative. These two situations directly contradict each other. For example, in the first case the negative mass test particle is gravitationally attracted toward the positive mass particle located at the origin, but in the second case the positive mass test particle is gravitationally repelled by the negative mass particle located at the origin. This contradiction makes equations (1) and (2) logically inconsistent if negative mass density were to exist. The only way to avoid this contradiction is to require that mass density be non-negative always. Requiring mass density $\rho_m$ be non-negative always is also consistent with the global symmetry transformations (39) through (41) where it was noted that the mass density field $\rho_m$ does not change sign under any of the symmetry transformations.

The properties of matter and antimatter and their associated phenomenology covered above are all derivative to (1) and (2). Additionally, matter-antimatter annihilation can be explained using the conservation of mass (25) which is also derivative to (1) and (2). Consider a region containing a mass current density $\rho_m u^\nu$ and another region containing a mass current density $\rho_m\left(-u^\nu\right)$ which because of the opposite sign of its four-velocity time component is a region of antimatter. If these two regions were to coincide at some point, then the net mass current density would vanish consistent with matter-antimatter annihilation, and the annihilation process would satisfy the mass conservation (25). Because these results are all derivative to (1) and (2), such matter-antimatter annihilation would have to be accompanied by the generation of electromagnetic energy as required by the conserved energy-momentum tensor, (2).

The symmetry transformation (40) has been employed here to illustrate matter-antimatter phenomenology using what is essentially an infinitesimal test-particle approach. Specifically, to understand the consequences of (1) and (2) an infinitesimal test particle interacting with an external electromagnetic or gravitational field was analyzed. One issue that was glossed over in this approach is that when considering a particular-like solution to (1) and (2) in which say the test particle is composed of matter, then it is not quite legitimate to use (40) to transform just the test particle to antimatter while leaving the external fields unchanged, this because the transformation (40) is a global symmetry transformation and would change the sign of the external electromagnetic field if applied everywhere. While the approach used here is generally acceptable when considering continuous fields in which legitimately infinitesimal test particles or regions can be considered because they are small enough to not perturb the external fields, it will be shown that the particle like-solutions to (1) and (2) are subject to quantization conditions which precludes such particles being treated as truly infinitesimal. Said another way, if infinitesimal test particles are not representative of the particle-like solutions to (1) and (2), then using transformation (40) to transform particle-like solutions only but not the external electromagnetic fields may not be quite legitimate. The upshot being, that when considering physical particle-like solutions to equations (1) and (2), the results derived in the forgoing section may not be exact.

## 3. SOLUTIONS TO EQUATIONS (1) AND (2)

To illustrate the theory that flows from equqtions (1) and (2), in this section I present three specific solutions. Comparisons are made between these solutions and the corresponding solutions to the classical M&EFEs noting both similarities between the two sets of solutions and where the solutions given by (1) and (2) go further than those of the classical M&EFEs.

### Spherically symmetric solution for the fields of a charged particle

Here a solution representing the asymptotic fields of a non-rotating, spherically symmetric, charged particle is investigated. It is demonstrated that the Reissner-Nordström metric with an appropriate choice for $a^{\lambda}$ and $F_{\mu\nu}$ satisfies (1) and (2) with $\rho_m$, $\rho_c$ and $u^{\lambda}$ being defined in terms of $a_{\lambda}$, $g_{\mu\nu}$ and $F_{\mu\nu}$ by (27), (19) and (20), respectively. To proceed, I draw on a solution for a spherically symmetric charged particle that was previously derived.[12] Starting with the Reissner-Nordström metric[13]

$$g_{\mu\nu} = \begin{pmatrix} \dfrac{1}{\left(1-\dfrac{2m}{r}+\dfrac{q^2}{r^2}\right)} & 0 & 0 & 0 \\ 0 & r^2 & 0 & 0 \\ 0 & 0 & r^2\mathrm{Sin}^2(\theta) & 0 \\ 0 & 0 & 0 & -\left(1-\dfrac{2m}{r}+\dfrac{q^2}{r^2}\right) \end{pmatrix} \tag{50}$$

and the Ricci tensor that follows from it

$$R_{\lambda}{}^{\nu} = \frac{q^2}{r^4}\begin{pmatrix} 1 & 0 & 0 & 0 \\ 0 & -1 & 0 & 0 \\ 0 & 0 & -1 & 0 \\ 0 & 0 & 0 & 1 \end{pmatrix} \tag{51}$$

I investigate a trial solution in which $a^{\lambda}$ is constant,

$$a^{\lambda} = \left(0, 0, 0, c_1\right) \tag{52}$$

with the value of the constant $c_1$ yet to be determined. Next, using the definition of charge current density from (12) and the expressions for $R^{\lambda}{}_{\nu}$ and $a^{\lambda}$ from (51) and (52), respectively, gives

$$\rho_c u^{\nu} = a^{\lambda} R_{\lambda}{}^{\nu} = \left(0, 0, 0, c_1 \frac{q^2}{r^4}\right) \tag{53}$$

Using the definitions for the charge density $\rho_c$ and the four-velocity $u^{\lambda}$ from (19) and (20), respectively, then gives,

$$\rho_c = \pm\left|c_1\right| \frac{q^2}{r^4} \sqrt{\left|1 - \frac{2m}{r} + \frac{q^2}{r^2}\right|} \tag{54}$$

and

$$u^{\lambda} = \left(0, 0, 0, \pm\frac{c_1}{\left|c_1\right|} \frac{1}{\sqrt{\left|1 - \frac{2m}{r} + \frac{q^2}{r^2}\right|}}\right) \tag{55}$$

The next step is to satisfy (1) by solving for $F_{\mu\nu}$. To simplify what follows I assume $q > m$ which allows me to drop the absolute values inside the square root in (54) and (55) as $1 - 2m/r + q^2/r^2$ is always greater than 0.[‡] Rather than tackling equation (1) head-on by directly solving the mixed system of first order partial differential equations that is (1), I instead solve the integrability equations (31), which are linear in $F_{\mu\nu}$ for $F_{\mu\nu}$. Proceeding in this manner, all the integrability equations are satisfied for $F_{\mu\nu}$ given by

$$F_{\mu\nu} = \begin{pmatrix} 0 & B_{\phi} & -B_{\theta} & E_r \\ -B_{\phi} & 0 & B_r & E_{\theta} \\ B_{\theta} & -B_r & 0 & E_{\phi} \\ -E_r & -E_{\theta} & -E_{\phi} & 0 \end{pmatrix} = \begin{pmatrix} 0 & 0 & 0 & \frac{(mr - q^2)}{r^3} c_1 \\ 0 & 0 & 0 & 0 \\ 0 & 0 & 0 & 0 \\ -\frac{(mr - q^2)}{r^3} c_1 & 0 & 0 & 0 \end{pmatrix} \tag{56}$$

‡ For example, consider an electron. In the geometric units used here the electron charge is $1.38\text{x}10^{-36}$ meters and the electron mass is $6.75\text{x}10^{-58}$ meters.

By direct substitution it is easily verified that $F_{\mu\nu}$ as given in (56) is indeed a solution of (1).[14] Choosing the value of the undetermined constant to be

$$c_1 = q/m \tag{57}$$

then gives an electric field that agrees with the Coulomb field for a point charge to leading order in $1/r$

$$F_{\mu\nu} = \begin{pmatrix} 0 & B_\phi & -B_\theta & E_r \\ -B_\phi & 0 & B_r & E_\theta \\ B_\theta & -B_r & 0 & E_\phi \\ -E_r & -E_\theta & -E_\phi & 0 \end{pmatrix} = \begin{pmatrix} 0 & 0 & 0 & \dfrac{q}{r^2} - \dfrac{q^3/m}{r^3} \\ 0 & 0 & 0 & 0 \\ 0 & 0 & 0 & 0 \\ -\dfrac{q}{r^2} + \dfrac{q^3/m}{r^3} & 0 & 0 & 0 \end{pmatrix} \tag{58}$$

Finally, the mass density $\rho_m$ is found using (27),

$$\rho_m = \frac{q^4}{m^2 r^4}\left(1 - \frac{2m}{r} + \frac{q^2}{r^2}\right) \tag{59}$$

To summarize, the following expressions for $g_{\mu\nu}$, $a^\lambda$, $F_{\mu\nu}$, $u^\lambda$, $\rho_c$ and $\rho_m$ comprise an exact solution to (1) and (2):

$$
\begin{gathered}
g_{\mu\nu} = \begin{pmatrix} \dfrac{1}{\left(1-\dfrac{2m}{r}+\dfrac{q^2}{r^2}\right)} & 0 & 0 & 0 \\ 0 & r^2 & 0 & 0 \\ 0 & 0 & r^2 \mathrm{Sin}^2(\theta) & 0 \\ 0 & 0 & 0 & -\left(1-\dfrac{2m}{r}+\dfrac{q^2}{r^2}\right) \end{pmatrix} \\
a^\lambda = s_{c-c} s_{m-a} \left(0,0,0,\frac{q}{m}\right) \\
F_{\mu\nu} = \begin{pmatrix} 0 & B_\phi & -B_\theta & E_r \\ -B_\phi & 0 & B_r & E_\theta \\ B_\theta & -B_r & 0 & E_\phi \\ -E_r & -E_\theta & -E_\phi & 0 \end{pmatrix} = s_{c-c} s_{m-a} \begin{pmatrix} 0 & 0 & 0 & \dfrac{q}{r^2}-\dfrac{q^3/m}{r^3} \\ 0 & 0 & 0 & 0 \\ 0 & 0 & 0 & 0 \\ -\dfrac{q}{r^2}+\dfrac{q^3/m}{r^3} & 0 & 0 & 0 \end{pmatrix} \\
u^\lambda = s_{m-a} \left(0,0,0,\frac{1}{\sqrt{1-\dfrac{2m}{r}+\dfrac{q^2}{r^2}}}\right) \\
\rho_c = s_{c-c} \frac{q^3}{m} \frac{\sqrt{1-\dfrac{2m}{r}+\dfrac{q^2}{r^2}}}{r^4} \\
\rho_m = \frac{q^4}{m^2} \frac{(1-\dfrac{2m}{r}+\dfrac{q^2}{r^2})}{r^4}
\end{gathered}
\qquad (60)
$$

In (60), the multiplicative parameters $s_{c-c}$ and $s_{m-a}$ in the equations for $a^\lambda$, $F_{\mu\nu}$, $u^\lambda$ and $\rho_c$ take on the values ±1 and correspond to the global charge-conjugation (39) and global matter-antimatter conjugation (40), respectively. Except for the possibility of both matter and antimatter solutions, the physical interpretation of solution (60) is almost identical to that of the classical M&EFEs, *i.e.*, a non-rotating, spherically symmetric particle having charge $q$ and mass $m$. The Reissner-Nordström metric tensor establishes that the theory based on (1) and (2) and Einstein's General Relativity predict the same gravitational fields in this case. However, solution (60) does differ from the classical picture in several ways. First, the mass and charge are not localized, with both $\rho_m$ and $\rho_c$ having a spatial extent that falls off as $1/r^4$. Next, the radial electric field

$$E_r = \frac{q}{r^2} - \frac{q^3/m}{r^3} = \frac{q}{r^2}\left(1 - \frac{q^2/m}{r}\right) \tag{61}$$

while agreeing with the Coulomb field $q/r^2$ to leading order in $1/r$ does have a higher order term. This next term depends on both the charge and mass of the particle. Taking an electron as an example, its electric field as given by (61) would be

$$E_r = \frac{q_e}{r^2}\left(1 - \frac{q_e^{\,2}/m_e}{r}\right) = \frac{q_e}{r^2}\left(1 - \frac{r_e}{r}\right) \tag{62}$$

where $r_e = q_e^{\,2}/m_e \sim 2.8x10^{-15}m$, the classical radius of an electron.

In the static particle-like solution to (1) and (2) just considered, the metric tensor $g_{\mu\nu}$ given by the Reissner-Nordström metric in (60) is expressed parametrically in terms of the particle's mass $m$ and charge $q$. Because the mass density $\rho_m$ and charge density $\rho_c$ are also specified as part of any solution to (1) and (2) as defined by (27) and (19), respectively, self-consistency boundary conditions exists in which the particle's total charge $q$ and total mass $m$ must agree with the spatially integrated charge and mass density, respectively. For the charge, this amounts to requiring,

$$q = \lim_{r\to\infty}\left(r^2 F_{14}\right) = \int \rho_c u^4 \sqrt{\gamma_{sp}}\, d^3x \tag{63}$$

where $\gamma_{sp}$ is the determinant of the spatial metric defined by[15]

$$\gamma_{sp\; i\,j} = g_{i\,j} - \frac{g_{i4}\, g_{j4}}{g_{44}} \tag{64}$$

and *i* and *j* run over the spatial dimensions 1, 2 and 3. For the mass, the analogous self-consistency boundary condition is

$$m = \lim_{r\to\infty}\left(r\,\frac{1+g_{44}}{2}\right) = \int \rho_m \left|u^4\right| \sqrt{\gamma_{sp}}\, d^3x \tag{65}$$

The reason for the absolute value of $u^4$ in the mass self-consistency condition (65) but not in the charge self-consistency condition (63) are the global symmetries (39) through (41) exhibited by the theory's equations (1) and (2), and the requirement that the self-consistency conditions exhibit those same symmetries. The conjecture being put forth here is that (63) and (65) represent self-consistency constraints on the charge and the mass, respectively, that any particle-like solution to (1) and (2) must satisfy if the solutions are to be physically realizable. Although not pursued further here, when considering metrics that include nonzero angular momentum, as for example would be required for particles having an intrinsic magnetic field, the same approach used here to quantize the particle's mass and charge could be used to quantize its angular momentum. Traditionally the quantization of mass, charge and angular momentum are introduced in quantum mechanical treatments but here are conjectured within the framework of a classical continuous field-theoretic description of nature and are another example of how the proposed theory departs from the classical M&EFEs.

For the spherically symmetric solution (60), the charge and mass self-consistency boundary conditions given by (63) and (65), respectively, both diverge on their RHS and so are not satisfied by the given solution. The upshot of this is that while (60) represents a formal mathematical solution of (1) and (2) that describe the asymptotic gravitational and electrical fields of a particle, it cannot represent a physically allowed solution. The possibility of finding solutions that satisfy both (1) and (2) and the charge and mass boundary conditions (63) and (65) remains an open question at this point. However, interesting possibilities exist beyond the spherically symmetric solution based on the Reissner-Nordström metric investigated here. For example, the modified Reissner-Nordström and modified Kerr-Newman metrics developed by S.M. Blinder[16] give finite values for the RHS of both (63) and (65).

The gravitational field predicted by the solution to (1) and (2) investigated here agrees with the corresponding solution to Einstein's General Relativity (34), both being described by the Reissner-Nordström metric. However, it is important to note that the classical General Relativity field equations (34) are not satisfied using the energy-momentum tensor of (2)

$$T^{\mu\nu} = \rho_m u^\mu u^\nu + F^\mu{}_\lambda F^{\nu\lambda} - \frac{1}{4} g^{\mu\nu} F^{\rho\sigma} F_{\rho\sigma} \tag{66}$$

with the values for $\rho_m$, $u^\nu$ and $F_{\mu\nu}$ as given in (60). However, Einstein's equation of General Relativity augmented by the $\Lambda^{\mu\nu}$ term on its RHS as given in (38) is trivially satisfied. For completeness, the values of $G^{\mu\nu}, T^{\mu\nu}$ and $\Lambda^{\mu\nu}$ that go with solution (60) are:

$$G^{\mu\nu}=\begin{pmatrix} \frac{q^2}{r^4}\left(1-\frac{2m}{r}+\frac{q^2}{r^2}\right) & 0 & 0 & 0 \\ 0 & -\frac{q^2}{r^6} & 0 & 0 \\ 0 & 0 & -\frac{q^2\csc^2(\theta)}{r^6} & 0 \\ 0 & 0 & 0 & -\frac{q^2}{r^4}\frac{1}{\left(1-\frac{2m}{r}+\frac{q^2}{r^2}\right)} \end{pmatrix}$$

$$T^{\mu\nu}=\begin{pmatrix} -\frac{q^2\left(q^2-mr\right)^2\left(q^2+r\left(-2m+r\right)\right)}{2m^2r^8} & 0 & 0 & 0 \\ 0 & \frac{q^2\left(q^2-mr\right)^2}{2m^2r^8} & 0 & 0 \\ 0 & 0 & \frac{q^2\left(q^2-mr\right)^2\csc^2(\theta)}{2m^2r^8} & 0 \\ 0 & 0 & 0 & \frac{3q^6+m^2q^2r^2+2q^4r\left(-3m+r\right)}{2m^2r^4\left(q^2+r\left(-2m+r\right)\right)} \end{pmatrix} \quad (67)$$

$$\Lambda^{\mu\nu}=G^{\mu\nu}+8\pi T^{\mu\nu}$$

In the context of classical General Relativity the interpretation of $\Lambda^{\mu\nu}$ is that of an energy-momentum tensor appropriate for dark matter and/or dark energy which serves as a source term for gravitational fields in addition to those sourced by $T^{\mu\nu}$, the energy-momentum tensor for ordinary matter and ordinary energy. However, in the context of the proposed geometricized theory of electrodynamics, $\Lambda^{\mu\nu}$ is expressed in terms of normal matter and normal energy through solutions to (1) and (2) which again emphasizes that the gravitation emerging from (1) and (2) differs from that of classical General Relativity. With questions today regarding the validity of classical General Relativity beyond the confines of our own solar system[17] and the inability to directly detect dark matter and dark energy, the interpretation of $\Lambda^{\mu\nu}$ in terms normal matter and normal energy in solutions to (1) and (2) is an interesting possibility.

**Radiative solutions for electromagnetic and gravitational waves**

Working in the weak field limit and using only equation (1) expressions are derived for a propagating electromagnetic plane wave in terms of the vector field $a^{\lambda}$ and the metric tensor $g_{\mu\nu}$. This example establishes a fundamental relationship between electromagnetic and gravitational radiation with both being manifestations of wave propagation of the underlying metric $g_{\mu\nu}$. To begin, consider an electromagnetic plane wave having frequency $\omega$, propagating in the +*z*-direction and polarized in the *x*-direction. The classical Maxwell tensor for this field is given by

$$F_{\mu\nu} = \begin{pmatrix} 0 & 0 & -B_y & E_x \\ 0 & 0 & 0 & 0 \\ B_y & 0 & 0 & 0 \\ -E_x & 0 & 0 & 0 \end{pmatrix} e^{i\omega(t-z)} \tag{68}$$

where $E_x$ and $B_y$ are the constant field amplitudes of the electromagnetic wave. Next, assume a near-Minkowski weak field metric given by

$$g_{\mu\nu} = \eta_{\mu\nu} + h_{\mu\nu} e^{i\omega(t-z)} \tag{69}$$

where $h_{\mu\nu}$ are complex constants satisfying $\left|h_{\mu\nu}\right| \ll 1$, $\eta_{\mu\nu} = diag[1,1,1,-1]$, and the vector field $a^\lambda$ is assumed to be constant and given by

$$a^\lambda = \left(a^1, a^2, a^3, a^4\right) \tag{70}$$

I proceed by substituting for $F_{\mu\nu}$ using (68), $g_{\mu\nu}$ using (69), and $a^\lambda$ using (70) into (1), and then only retaining terms to first order in the fields $h_{\mu\nu}$ and $F_{\mu\nu}$, both of which are assumed to be small and of the same order.[18] Doing this leads to a set of 8 independent linear equations for the 16 unknown constants: $h_{\mu\nu}$, $a^\lambda$, $E_x$ and $B_y$. Solving these 8 independent equations gives 8 the field components $E_x$, $B_y$, $h_{13}$, $h_{22}$, $h_{23}$, $h_{34}$ , $a^2$ and $a^3$ in terms of 8 free parameters $a^1$, $a^4$, $h_{11}$, $h_{12}$ , $h_{14}$ , $h_{24}$, $h_{33}$, and $h_{44}$ :

$$\begin{aligned} E_x &= i\,\omega \frac{\left(h_{11}{}^2 + h_{12}{}^2\right)}{2\,h_{11}} a^1 \\ B_y = E_x &= i\,\omega \frac{\left(h_{11}{}^2 + h_{12}{}^2\right)}{2\,h_{11}} a^1 \end{aligned} \tag{71}$$

$$g_{\mu\nu} = \eta_{\mu\nu} + \begin{pmatrix} h_{11} & h_{12} & -h_{14} & h_{14} \\ h_{12} & -h_{11} & -h_{24} & h_{24} \\ -h_{14} & -h_{24} & h_{33} & -\frac{1}{2}\left(h_{33}+h_{44}\right) \\ h_{14} & h_{24} & -\frac{1}{2}\left(h_{33}+h_{44}\right) & h_{44} \end{pmatrix} e^{i\omega(t-z)} \tag{72}$$

and

$$a^{\lambda} = \left( a^1, a^1 \frac{h_{12}}{h_{11}}, a^4, a^4 \right) \tag{73}$$

This solution illustrates several ways in which the new theory departs from the classical physics view of electromagnetic radiation. Of most significance, the undulations in the electromagnetic field are due to undulations in the underlying metric field $g_{\mu\nu}$ given in (72). This result also underscores that the existence of electromagnetic radiation is forbidden in strictly flat space-time. An interesting aspect of this solution is that while the electromagnetic field necessitates the presence of an underlying gravitational radiation field, the underlying gravitational field is not completely defined by the electromagnetic field. The supporting gravitational radiation has 6 undetermined constants $\left(h_{11}, h_{12}, h_{14}, h_{24}, h_{33}, h_{44}\right)$ with the only restriction being $\left|h_{\mu\nu}\right| \ll 1$ and $h_{11} \neq 0$ as required by (71). Further insight into the physical content of the metric (72) is evident after making the infinitesimal coordinate transformation from $x^{\mu} \rightarrow x'^{\mu}$ defined by

$$\begin{pmatrix} x \\ y \\ z \\ t \end{pmatrix} \rightarrow \begin{pmatrix} x' \\ y' \\ z' \\ t' \end{pmatrix} = \begin{pmatrix} x - \frac{i}{\omega} h_{14}\, e^{i\omega(t-z)} \\ y - \frac{i}{\omega} h_{24}\, e^{i\omega(t-z)} \\ z + \frac{i}{2\omega} h_{33}\, e^{i\omega(t-z)} \\ t + \frac{i}{2\omega} h_{44}\, e^{i\omega(t-z)} \end{pmatrix} \tag{74}$$

and only retaining terms to first order in the *h's*. Doing this, the metric (72) is transformed to

$$g'_{\mu\nu} = \eta_{\mu\nu} + \begin{pmatrix} h_{11} & h_{12} & 0 & 0 \\ h_{12} & -h_{11} & 0 & 0 \\ 0 & 0 & 0 & 0 \\ 0 & 0 & 0 & 0 \end{pmatrix} e^{i\omega(t-z)} \tag{75}$$

while $E'_x$ and $B'_y$, the transformed electric and magnetic field amplitudes, respectively, are identical to $E_x$ and $B_y$ given in (71). Note, only the $h_{11}$ and $h_{12}$ components of the metric (75) have an absolute physical significance and $h_{22} = -h_{11}$, which makes the gravitational plane wave solution (75) identical to the gravitational plane wave solution of the classical Einstein field equations.[19], [20]

The forgoing analysis demonstrated the necessity of having an underlying gravitational wave to support the presence of an electromagnetic wave, but the converse is not true and gravitational radiation can exist independent of electromagnetic radiation. The following analysis demonstrates this by solving for the structure of gravitational radiation in the absence of electromagnetic radiation. Following the same weak field formalism for the unknown fields $h_{\mu\nu}$ given in (69), but this time zeroing out $E_x$ and $B_y$ in (68), leads to the following solutions for $g_{\mu\nu}$ and $a^\lambda$

$$g_{\mu\nu} = \eta_{\mu\nu} + \begin{pmatrix} h_{11} & h_{12} & -h_{14} & h_{14} \\ h_{12} & \dfrac{h_{12}{}^2}{h_{11}} & -h_{24} & h_{24} \\ -h_{14} & -h_{24} & h_{33} & -\dfrac{h_{33}+h_{44}}{2} \\ h_{14} & h_{24} & -\dfrac{h_{33}+h_{44}}{2} & h_{44} \end{pmatrix} e^{i\omega(t-z)} \tag{76}$$

and,

$$a^\lambda = \left( a^1, -a^1 \frac{h_{11}}{h_{12}}, a^4, a^4 \right) \tag{77}$$

Both $g_{\mu\nu}$ given by (76) and $a^\lambda$ given by (77) are modified from their solutions in the presence of an electromagnetic wave as given by (72) and (73), respectively. Performing a transformation to the same primed coordinate system as given in (74), here gives the metric field

$$g'_{\mu\nu} = \eta_{\mu\nu} + \begin{pmatrix} h_{11} & h_{12} & 0 & 0 \\ h_{12} & \dfrac{h_{12}{}^2}{h_{11}} & 0 & 0 \\ 0 & 0 & 0 & 0 \\ 0 & 0 & 0 & 0 \end{pmatrix} e^{i\omega(t-z)} \tag{78}$$

illustrating again that only the $h_{11}$ and $h_{12}$ components have an absolute physical significance. Of particular note is the change in the value of the $h_{22}$ component depending on whether the gravitational wave supports an electromagnetic wave as in (75) or is standalone as in (78).

In summary, equation (1) has been shown to have two distinct classes of radiative solutions, one in which gravitational and electromagnetic radiation are coupled, and the other in which there is only gravitational radiation. In the solution with the coupled gravitational and electromagnetic radiation, the gravitational wave is identical to the weak field gravitational wave of classical General Relativity. Here again, the solutions of (1) are seen to be consistent with those of the classical M&EFEs but to then go further by providing an underlying unification between electromagnetic and gravitational phenomena. Finally, because both gravitational and electromagnetic radiation fields are due to undulations of the metric field $g_{\mu\nu}$ their propagation speeds are predicted to be identical. This prediction continues to be refined experimentally. For example, observations made during the binary neutron star merger in NGC 4993, 130 million light years from Earth.[21] The nearly simultaneous detection, within 2 seconds of each other, of gravity waves[22] and a burst of gamma rays[23] from this event experimentally constrain the propagation speed of electromagnetic and gravitational radiation to be the same to better than 1 part in $10^{15}$.

**Solution having a maximally symmetric 3-dimensional subspace**

Next, I consider the time-dependent Friedmann–Lemaître–Robertson–Walker (FLRW) metric

$$g_{\mu\nu} = \begin{pmatrix} \frac{R_s^2(t)}{1-kr^2} & 0 & 0 & 0 \\ 0 & R_s^2(t)r^2 & 0 & 0 \\ 0 & 0 & R_s^2(t)r^2\,\mathrm{Sin}^2(\theta) & 0 \\ 0 & 0 & 0 & -1 \end{pmatrix} \tag{79}$$

where $k$ equals +1, 0 or -1 depending on whether the spatial curvature is positive, zero or negative, respectively, and $R_s(t)$ is a time-dependent scale factor. Just as in the case of General Relativity where the FLRW metric is a cosmological solution representing a homogeneous and isotropic universe it is the same for equation (1) with an appropriate choice for the time development of $R_s(t)$. To derive the time dependence of $R_s(t)$ I note the 3-dimensional spatial subspace of (79) is maximally symmetric and so any tensor fields that inhabit that subspace must also be maximally symmetric.[24] Specifically, this restricts the form of $a^{\mu}$ to be

$$a^{\mu} = \left(0, 0, 0, a^{4}\left(t\right)\right) \tag{80}$$

and forces the antisymmetric Maxwell tensor to vanish,

$$F_{\mu\nu} = 0 \tag{81}$$

Because $F_{\mu\nu}$ vanishes so must $F_{\mu\nu;\kappa}$,

$$F_{\mu\nu;\kappa} = 0 \tag{82}$$

which on substitution in (1) forces,

$$a^{\lambda} R_{\lambda\kappa\mu\nu} = 0 \tag{83}$$

This in turn forces,

$$a^{\lambda} R_{\lambda}^{\ \nu} = 0 \tag{84}$$

which gives $\rho_c = 0$ by equation (19). Substituting $a^{\mu}$ given by (80), and the FLRW metric given by (79) into (83) then leads to the following set of equations to be satisfied:

$$\begin{aligned}
a^4(t)R_{4114} &= a^4(t)\left(\frac{R_s(t)}{k\,r^2 - 1}\frac{d^2R_s(t)}{dt^2}\right) = 0 \\
a^4(t)R_{4224} &= a^4(t)\left(-r^2\,R_s(t)\frac{d^2R_s(t)}{dt^2}\right) = 0 \\
a^4(t)R_{4334} &= a^4(t)\left(-r^2\,R_s(t)\,\mathrm{Sin}^2(\theta)\frac{d^2R_s(t)}{dt^2}\right) = 0
\end{aligned} \tag{85}$$

with all other components of (83) not listed in (85) being trivially satisfied, *i.e.*, $0 = 0$ . The nontrivial components given in equations (85) are all satisfied if

$$\frac{d^2R_s(t)}{dt^2} = 0 \tag{86}$$

or,

$$R_s(t) = R_{s\,0} + v_s\,t \tag{87}$$

where $R_{s\,0}$ is the scale factor at time *t=0*, and $v_s$ is its constant rate of change.

Summarizing, the predictions of equation (1) for a homogeneous and isotropic solution are:

1. It must be charge neutral, $\rho_c = 0$.
2. The scale factor $R_s(t)$ changes linearly with time.
3. The spatial curvature of the solution can be positive, negative or 0.

The second prediction above regarding the linear time dependence of the scale factor $R_s(t)$ differs from the predictions of the Friedmann models of General Relativity and again emphasizes that the theory of gravitation emerging here differs from that described by classical General Relativity.

**A note on superluminal transport**

All results presented in this manuscript have been mathematically derivative to equations (1) and (2) which serve as the axioms for the developed theory. One development that was noted but not covered further is the prediction by equation (20) of superluminal transport if $a^{\lambda}R_{\lambda}{}^{\nu}$ is space-like. Because the developed theory flows axiomatically from (1) and (2), and because there is nothing *a priori* that precludes the possibility of $a^{\lambda}R_{\lambda}{}^{\nu}$ being space-like I have carried this as a possibility, although one that must be regarded as speculative at this point because the specific solutions investigated in this section have not exhibited it. Although not pursued further here, the possibility of superluminal transport in the context of a classical field theory may be an interesting and timely avenue of investigation as recent research has suggested the possible existence of nonlocal correlations stronger than those predicted by quantum theory.[25]

## 4. CONCLUSION

Using two equations, equation (1) which is new and equation (2) which is an already well-established foundation of classical physics, a continuous field theory of electromagnetism is developed from which gravitation and antimatter emerge in solutions. The choice of equation (1) as the foundation for the geometricized theory of electromagnetism developed here was driven by the goal to preserve the classical Maxwell equations. This is accomplished using a vector field $a^{\lambda}$ to couple the Maxwell tensor to the R-C tensor in (1) such that Maxwell's homogeneous and inhomogeneous equations are derivative to (1) but with their classical interpretations furthered by the imposed definitions on the charge density $\rho_c$ and the four-velocity $u^{\lambda}$ in terms of $a^{\lambda}$ and $R_{\lambda}{}^{\kappa}$. Next, mass density $\rho_m$ is introduced through the conserved energy-momentum tensor (2). Using equation (2) and the already derived Maxwell equations, the conservation of mass equation and the Lorentz force law are derived. Using the Lorentz force law, the mass density $\rho_m$ is then defined geometrically. The commonality of the definitions for $\rho_c$, $\rho_m$ and $u^\lambda$ in terms of $a^\lambda$ and $R_\lambda^{\ \kappa}$ interrelates these quantities and advances their interpretations beyond what they are in classical electrodynamics and ultimately leads to the emergence of gravity and antimatter in solutions to (1) and (2). The gravity that does emerge is equivalent to Einstein's field equation of General Relativity augmented by a symmetric and conserved tensor field $\Lambda^{\mu\nu}$. In the context of General Relativity, $\Lambda^{\mu\nu}$ mimics the properties of dark matter and/or dark energy, *i.e.*, it is a source of gravitational fields but exhibits no other interactions in the theory. Using the global symmetries of equations (1) and (2), the emergence of antimatter and its behavior in electromagnetic and gravitational fields is then developed. In external gravitational fields, matter and antimatter are shown to have identical responses, both being attracted by the source of the gravitational field, while antigravity is shown to be inconsistent with solutions to (1) and (2) and so does not exist. In external electromagnetic fields, matter and antimatter are shown to have responses consistent with those of classical electromagnetic theory. Finally, using specific solutions to (1) and (2), features of the geometricized theory that go beyond the classical M&EFEs are illustrated. As examples, a mechanism for quantizing static particle-like solutions is proposed, and the unification of electromagnetic and gravitational waves is demonstrated.

## Disclaimer

The work presented within was reported in a preliminary form in references [2] and [26]. The coupling between the Maxwell tensor and the R-C tensor given in equation (1) was first reported in those references, although in a somewhat modified form. New to this manuscript is the discussion of the global symmetries of equations (1) and (2), and based on those global symmetries the interpretation of the particle-like solution has been advanced as has the discussion of the self-consistency conditions and the quantization of stationary particle-like solutions. The discussion of Einstein's equation of General Relativity augmented by a term that can mimic the properties of dark matter and/or dark energy is new to this manuscript, as is the discussion of the solution based on the FLRW metric. The present manuscript also corrects an error in the weak field analysis of reference [2] leading to an expanded discussion of electromagnetic and gravitational radiation. The discussion on the impossibility of antigravity is new as is the speculation on superluminal transport if $a^\lambda R_\lambda^{\ \nu}$ is space-like.

## Acknowledgements

For his penetrating insights and suggestions offered during the early phases of this work[26] I would like to express my appreciation and thanks to David Eimerl. For many useful conversations and critical comments during the preparation of this manuscript I would like to thank my colleagues at Lawrence Livermore National Laboratory: Steve Libby, Michael Schneider, Mark Rotter, Ed Hartouni, Charles Boley and Alexander Rubenchik.

This work was performed under the auspices of the U.S. Department of Energy by Lawrence Livermore National Laboratory under Contract DE-AC52-07NA27344.

---

[26] Beach, Raymond J. "A Representation of the Electromagnetic Field in the Presence of Curvature." *Physics Essays* 12 (1999): 457.